# A Demonstration of Spectral and Spatial Interferometry at THz Frequencies

**William F. Grainger**[1,*], **Roser Juanola-Parramon**[2], **Peter A. R. Ade**[1], **Matt Griffin**[1], **Flo Liggins**[1], **Enzo Pascale**[1], **Giorgio Savini**[2], **Bruce Swinyard**[2]

[1]*Astronomy Instrumentation Group, School of Physics and Astronomy, Cardiff University, Cardiff*
[2]*Optical Science Laboratory, Department of Physics and Astronomy, University College London, London*

*Corresponding Author: William.grainger@astro.cf.ac.uk*

## Abstract

A laboratory prototype spectral/spatial interferometer has been constructed to demonstrate the feasibility of the double Fourier technique at Far Infrared (FIR) wavelengths (0.15 - 1 THz). It is planned to use this demonstrator to investigate and validate important design features and data processing methods for future astronomical FIR interferometer instruments. In building this prototype we have had to address several key technologies to provide an end-end system demonstration of this double Fourier interferometer. We report on the first results taken when viewing single slit and double slit sources at the focus of a large collimator used to simulate real sources at infinity. The performance of the prototype instrument for these specific field geometries is analyzed to compare with the observed interferometric fringes and to demonstrate image reconstruction capabilities.

## 1  Introduction

In recent years, far infrared (FIR) observatories such as *Spitzer* and *Herschel* have made great advances in galactic and extragalactic astronomy, but are fundamentally limited by aperture size (*Spitzer*) and thermal background (*Herschel*). Future FIR missions are now being studied which can provide major improvements in sensitivity through active cooling of large apertures (*SPICA*) and spatial resolution through the use of interferometry (*SPIRIT*). These instruments would enable in-depth investigations in spectral and spatial detail of the material and processes involved in the formation of galaxies and proto-planetary systems. The main science drivers are dependent on obtaining spatial resolutions of ≤ 0.2 arc seconds which are only achievable with an interferometer. With this improvement we could resolve structures ~1AU in planetary systems at distances within 50pc; understand the evolution of massive black holes with their host galaxies; trace the assembly of Milky Way type galaxies at higher reshifts (z>6); resolve the Cosmic Infrared Background (CIRB) into individual sources and study their internal structure, physics and chemistry; and observe the formation of the first stars from primordial material. These science objectives also require ultra-sensitive detectors and the modest spectral capability provided by this double Fourier technique (resolving power ~ 1000).

The scientific case and technical challenges of space-borne long-baseline FIR interferometry have been studied in the USA [1] and in Europe [2] involving tethered or free-flying 3-m class cold telescopes with adjustable spacing for good *u-v* plane coverage. Theoretical studies and laboratory demonstrations have also been reported[3, 4]. Equipped with sensitive superconducting detector arrays, such a facility would enable spatially-resolved photometry and spectroscopy of high-redshift galaxies and resolve the detailed structure of planetary systems in the process of forming. As an intermediate step, SPIRIT, a shorter-baseline interferometer (several tens of metres) with boom-mounted apertures has been studied [5]. The implementation of a space-borne facility poses many technical challenges which need to be addressed before a robust design can be developed. Two balloon-borne FIR interferometer pathfinders are now in development. FITE [6], the Far Infrared Interferometric Experiment, is a Fizeau interferometer with an 8-m baseline giving 4" resolution at 150 μm, and is expected to have its first flight in 2012. BETTII, the Balloon Experimental Twin Telescope for Infrared Interferometry [7] is based on a double (spatial and spectral) Fourier interferometer similar to the SPIRIT concept with a first flight expected within the next five years. BETTII will operate in the wavelength range 30-50 and 60-90 μm. These balloon experiments, together with continuing laboratory-based FIR technology development and advances in detector arrays, will pave the way for a future FIR interferometer space mission.

In the proposals listed above many of the technical challenges have been identified: high sensitivity detectors (NEPs $\leq 10^{-19}$ W Hz$^{-1/2}$), cooled apertures, and appropriate data processing algorithms. To address a number of the instrumental issues we have constructed a laboratory prototype instrument which can cover the FIR spectral range. Operation of this prototype system will allow several key technologies such as efficient FIR beam combination, optical design, quasioptical dichroics, beam dividers, and polarizers to be investigated and characterized. In addition, the demonstration of complete end-end system performance will allow optimization of data processing algorithms and 3-D spatial-spectral reconstruction. The system incorporates a large collimator, making it possible to achieve fringe detection from artificial sources and thus to develop data processing techniques to enable source reconstruction to be demonstrated. In this paper we report on the successful observation of spectral/spatial fringes and show how the source morphology can be reconstructed in a few well defined cases.

## 2 Basic Principles

Here we present a basic mathematical approach to illustrate the nature of the fringe patterns that we expect to observe. The interferometer is designed to resolve the light from an extended scene spectrally and spatially. We start here with a description of a basic two beam spatial interferometer viewing a quasi-monochromatic point-like source, before broadening the discussion to generalized sources and then, finally, to the spectral domain. For any basic two beam interference experiment in which quasi-monochromatic light from a point-like source is sent along two separate paths of varying delay before recombining, it can be shown that the observed output intensity *I* is given by (e.g. [8]):

$$I = I_1 + I_2 + 2\sqrt{I_1}\sqrt{I_2}\,|\gamma_{12}|\cos(\beta_{12} - \delta), \quad (1)$$

where $I_1$ and $I_2$ are the intensities received from each interferometer arm (1 and 2 respectively), $\delta = 2\pi(x_2 - x_1)/\lambda$ is the phase delay between the interferometer arms and $x_1$ and $x_2$ are the optical path lengths of the two arms. The third term of equation (1) is the result of the cross correlation between the time averaged electric field vectors superimposed from the two interferometer arms which manifest themselves as the intensity interference modulation as the spatial delay is varied. Following the standard convention we use the mutual coherence function relating the emission from two points on the source such that:

$$\gamma_{12} = |\gamma_{12}| exp(\beta_{12}) \tag{2}$$

where $\gamma_{12}$ is also known as the complex degree of coherence for the emission which will vary from being correlated for neighboring points to uncorrelated for more distant points of emission on the same source. The magnitude of the complex degree of coherence, $|\gamma_{12}|$, will vary from unity (fully correlated) for neighboring points to zero (completely uncorrelated) for more distant points of emission on the same source. The phase, $\beta_{12}$, is dependent on systematic differences in the wave propagation within the interferometer arms which is in addition to the baseline created phase delay, $\delta$. It is easy to see that for coherent emission from a single source point, equation (1) reduces to a Young's slit type cosine fringe pattern if $I_1 = I_2$ and $\gamma_{12} = 1$. In a more realistic case, we need to cope with a varying degree of coherence for discrete sources and totally uncorrelated emission from other sources within the same interferometer beam. For a single partially coherent source, this involves integrating the mutual coherence over the source area as is shown in standard optics texts (e.g., [8,9]) resulting in the van Cittert-Zernike theorem. This states that the phase coherence factor ($\gamma_{12}$) for two points on a plane illuminated by a parallel plane quasi-monochromatic self luminous surface is equal to the normalised complex amplitude at one of the points in the Fraunhofer diffraction pattern associated with a spherical wave centred on the other point when the wave passes through an aperture whose boundary is the same as that of the source. The variation of amplitude across the wave is the same as the variation of luminance across the source. It is important to note here that the Fourier transform of the aperture field gives the amplitude of the diffraction field which conveniently provides the reciprocity between the measured fringes and the source morphology (e.g., [9]).

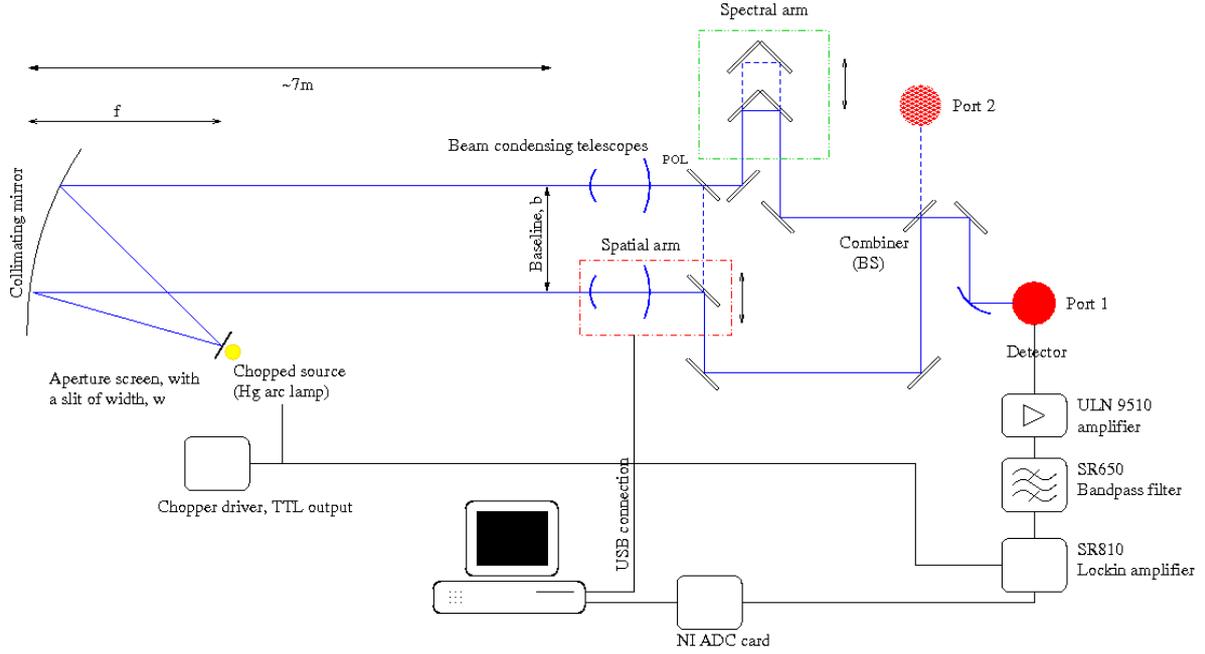

**Figure 1: Schematic of the interferometer. The dash-dot-dot box shows the moving part for the spectral arm, and dash-dot box shows the moving parts for the spatial arm. POL indicates the location of a polarizing grid added to turn the dual telescope (port) system into a single port system with the spatial arm removed. Port 1 and Port 2 represent the two complementary output ports of the interferometer; we only use Port 1**

Figure 1 shows a schematic of a laboratory implementation of a two element spectral/spatial interferometer. The laboratory instrument comprises two main parts; a source simulator and the spectral-spatial interferometer. The source simulator uses a large collimating mirror of focal length $f$ which produces a flat wave-front at the two input telescopes of the interferometer, simulating what would be received from a real astronomical source at infinity. Different apertures can be placed in front of the source. The spectral/spatial interferometer has two antennas, of diameter $D_{ant}$, separated by a baseline $b$.

For a single slit source, of width $w$, the complex degree of coherence is given by

$$\gamma_{12} = sinc(v) \qquad (3)$$

where $\beta_{12}$ is the phase of the coherence function, and $v$ is given by

$$v = \frac{2\pi b \theta_{source}}{\lambda}, \qquad (4)$$

with
$$\theta_{source} = \frac{w}{f} \qquad (5)$$

The quasi-monochromatic intensity for a single wide ($w > \lambda$) slit at the output of an interferometer using two small circular antennas of diameter $D_{ant}$ is then:

$$I(\phi,b) = \left(\frac{2J_1(u)}{u}\right)^2 [1 + \text{sinc}(v)\cos(\beta_{12}(v) - \delta)], \quad (6)$$

where the argument of the Bessel function is $u = (2\pi D_{ant}\phi)/\lambda$, the interferometer phase delay is $\delta = (2\pi b \sin\phi)/\lambda$, and $\phi$ is the angle of the source with respect to the nominal zero-difference delay position (sometimes known as the phase center in a spatial interferometer), which, in this case, is assumed to be same as the pointing center of the telescope apertures.

The other point to note that is that in the case of incoherent detection we measure intensity which is always positive, so the phase of the complex degree of coherence, $\beta_{12}(v)$, must be such that:

$$\begin{aligned}\beta_{12}(v) &= 0 \quad \text{when} \quad \text{sinc}(v) > 0 \\ &= \pi \quad \text{when} \quad \text{sinc}(v) < 0\end{aligned}. \quad (7)$$

We can determine the visibility of the fringes by observing that the maximum and minimum of equation (7) are:

$$\begin{aligned}I_{max}(\phi,b) &= \left(\frac{2J_1(u)}{u}\right)^2 (1 + |\text{sinc}(v)|) \\ I_{min}(\phi,b) &= \left(\frac{2J_1(u)}{u}\right)^2 (1 - |\text{sinc}(v)|)\end{aligned}, \quad (8)$$

giving the fringe visibility as

$$V(\phi,b) = |\text{sinc}(v)|. \quad (9)$$

A useful criterion for the conditions required to observe fringes clearly can be determined by setting this function to a specific value. It is usual to set $v = 1$ which gives $V(\phi,b) = 0.84$ as a reasonable criterion [8]. This then fixes the allowable source size, wavelength and baseline for an interference experiment.

The derivation of Equation (7) was outlined for a single slit. Generalizing this, it can be seen that the complex degree of coherence represents the Fourier transform of the actual source(s) brightness distribution for that specific wavelength. In general, we have:

$$I(\phi,b) = \left(\frac{2J_1(u)}{u}\right)^2 (1 + |\gamma_{12}|\cos(\beta_{12} - \delta)), \quad (10)$$

where $\gamma_{12}$ is, as before, the Fourier transform of the source(s) intensity distribution.

We mention here two further analytical examples for comparison with results in later sections: two equal slits of width $w$ and separation $d$, and two slits of equal width but transmission ratio $t$. In case of slits of equal size and equal transmission, the complex degree of coherence function is

$$\gamma_{double,12} = |\text{sinc}(v)\cos(v_{double})|\exp(\beta_{double,12} - \delta) , \qquad (11)$$

where $\beta_{double,12}$ is defined similarly to $\beta_{12}$ in equation (8), and

$$v_{double} = \frac{\pi b d}{\lambda f}, \qquad (12)$$

where $f$ is, as before, the focal length of collimator.

For slits of equal width and with transmission ratio $t$, the complex coherence function is

$$\gamma_{uneq,12} = |\cos(v) - (1-t).[\cos(v) + i.\sin(v)]|\exp(\beta_{uneq,12} - \delta) \qquad (13)$$

In general, for any sky scene we need to sample the source brightness distribution function as widely as possible in the Fourier spatial frequency domain (*u-v* plane) so that we can produce an accurate intensity distribution for the scene. With the addition of a spectral delay, we can record the fringe intensity for a particular baseline over a range of frequencies simultaneously, which makes this technique very attractive. The spectral optical path difference, $\Delta$, is easy to incorporate into the interferometer design, and, for the case of a single slit, simply adds into Equation (7) an additional phase factor of $\delta_{spect} = 2\pi\sigma\Delta$ , where $\sigma$ is the wavenumber, such that

$$I(\phi,b) = \left(\frac{2J_1(u)}{u}\right)^2 \left(1 + |\text{sinc}(v)|\cos(\beta_{12}(v) - \delta - \delta_{spect})\right). \qquad (14)$$

Whilst recording the spectral data, the baseline is fixed and the source is at a fixed angle with respect to the phase center of the interferometer. The modulation arises only from the change in $\delta_{spect}$ as the mirror is scanned. The spectral intensity, $S(\sigma)$, is however modified by the spectral dependence inherent in $v = 2\pi\sigma\theta_{source}b$ and $u = 2\pi\sigma D_{ant}\phi$. For the band of wavenumber covered ($0 - \sigma_{max}$), the spectral fringe intensity from the single slit source is thus given by:

$$I(\Delta) = \int_0^{\sigma_{max}} S(\sigma)\left(\frac{2J_1(u)}{u}\right)^2 \left(1 + |\text{sinc}(v)|\cos(\beta_{12}(v) - \delta - \delta_{spect})\right)d\sigma \qquad (15)$$

The form of equation (16) is identical to that derived for a polychromatic source illuminating a Michelson spectral interferometer where application of the Fourier integral theorem allows extraction of the spectral intensity for each interferometer baseline. We refer the reader here to standard texts on the extraction of the spectral data [10].

## 3  The laboratory demonstrator

We have built, aligned and operated a spectral spatial interferometer, based on the schematic shown in Figure 1. We chose to operate in the waveband 5 to 35 cm$^{-1}$ to provide a complement to the suggested BETTII bands, to build on our experience in this waveband with FTS instruments, and because the longer wavelengths allows for a less stringent tolerance. The laboratory testbed, unlike an observatory instrument, requires two parts; a source simulator and the spectral-spatial interferometer. In this section we describe both of these parts in detail.

## 3.1 Source and collimation unit

The source is a water-cooled mercury (Hg) arc lamp which has an equivalent blackbody-like emission at millimeter and submillimeter wavelengths of near 2000 K. Its emission is limited by a 10 mm diameter aperture in its housing. Initially we choose to use a step-and-integrate technique for recording the spectral fringe data. It was therefore necessary to provide amplitude modulation of the source to overcome the low-frequency noise arising from fluctuations in the room emission and in the detector unit. A taut-band modulator, similar to commercial designs [11], with a chopped aperture of 16 mm diameter, oscillating at 28 Hz, was therefore placed close to the Hg arc lamp. The scene aperture, constructed of aluminum tape across a stainless steel ring, was placed directly in front of the modulator blade at the focus of the collimator as shown in Figure 2. The scene comprised two nearly identical slits, one of height 6.9 mm and the other of 7.1 mm. Examination with an optical microscope showed the first slit is $1.01^{+0.03}_{-0.05}$ mm wide, the second $1.14^{+0.14}_{-0.20}$ mm, with the centers separated by $5.50^{+0.09}_{-0.06}$ mm. This geometry was considered to be close to the ideal case of two identical parallel slits given that the shortest wavelength was 0.3mm.

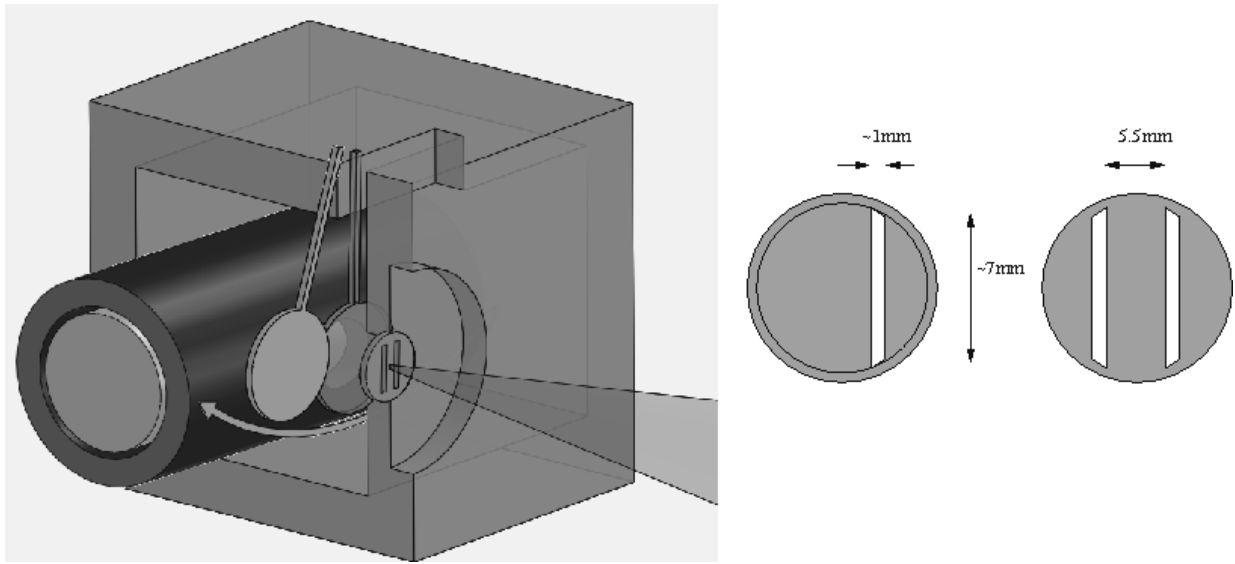

**Figure 2: Left: Detail of the Hg arc lamp, taut-beam chopper and slit holder. Right: schematics of the two slit configurations used.**

The collimator mirror is a 1-m radius gold-plated carbon-fiber section from the 6-segment BLAST telescope ([12]). It has a spherical surface of focal length 2.1 m from the mirror pole and generates a flat wavefront for a near on-axis source at the two interferometer telescope inputs. This system then has a focal plane scale of 1.64 arcmin/mm. The maximum interferometer baseline achievable before either of the beams from the input telescopes (see below) walk-off the collimator is 400 mm which corresponds to a maximum spatial resolution of 8.59 arcmin at 1 mm wavelength or 3.01 arcmin at 350 µm.

Initially we used one of the two slits (by covering the other with aluminum tape) to mimic a point-like source, noting that using our interference criterion in Section 2 it would just start to be resolved at the shortest wavelength at maximum baseline, resulting in a loss in fringe visibility. To observe spatial fringes clearly, we then uncovered the second slit parallel to this to simulate a double source. As distance between the centers of the two slits was 5.5 mm (9.00 arcmin), they should be easily resolvable at the shortest wavelength. We also had the option of covering the second slit with a low-pass filter (edge at 21cm$^{-1}$) to demonstrate extraction of spectral and spatial fringe data. As the interferometer is only sensitive in one spatial dimension, the slits were arranged perpendicular to the projected scanning direction to increase the available signal.

### 3.2 Interferometer

The light path in the interferometer is shown in Figure 1. The two input telescopes, with 100 mm diameter apertures and 25 mm diameter secondaries, are configured as 3:1 beam condensers. Note that this corresponds to an under-illumination of the primary collector. All other mirrors on the optical bench are 50 mm square optical quality flats except the final parabolic condenser which has a focal length of 304 mm. To achieve different spatial baselines, one of the input telescopes and its associated beam steering mirror are mounted on a Thorlabs Long Travel Stage (see box labeled "spatial arm" in Figure 1) with a maximum travel of 300 mm. At their closest separation the two telescope units are nearly touching so the minimum baseline is ~ 100 mm. The maximum spatial separation achievable is then ~ 400 mm, matching the available travel before this beam walks-off the collimator. Hereafter this is referred to as the "spatial arm" of the interferometer.

Having initially decided on using a step and integrate method for the Fourier Transform spectrometer and thus having less stringent motion control requirements than for a fast-scanning system, we used a second Thorlabs translation stage to provide a maximum path difference of 300 mm for the fold mirrors as shown by the boxed area labeled "Spectral arm" in Figure 1. Hereafter we will refer to this as the "spectral arm" of the interferometer. In the optical configuration as shown it should be noted that as the spatial baseline is changed, the position of the Zero Path Difference (ZPD) for the spectral arm shifts by half of the baseline change to balance the two optical paths to the beam divider from an assumed on-axis source. We will return to this point later as the notion that there is a fixed ZPD is incorrect for an imaging spatial interferometer. However, to understand the initial data sets it is useful to retain this notion here. To observe a complete set of spectral/spatial fringes requires that for a fixed baseline we scan the spectral arm to acquire the appropriate (spatially modulated) spectral data, then move both the baseline and the starting point of the FTS mirror and take the next set of (spatially modulated) spectral fringe data. Thus, as the interferometer baseline increases, the location of the ZPD (for an on-axis single point-like source) walks down the spectral arm. The spectral arm is thus also working as a path compensator.

The spectral fringe data are acquired every 16 µm of travel, or 32 µm of optical path difference (OPD), which corresponds to Nyquist sampling at a frequency of 4.7 THz. The beams from the two arms of the interferometer are combined at a 50:50 metal-mesh intensity beam splitter of 65 mm diameter which covers the spectral band from 5 to 33 cm$^{-1}$ (0.15 - 1 THz) [13]. As the maximum frequency is ~1 THz (see below), the data are therefore over-sampled by a factor of ~ 5 which enables accurate phase correction. For the data sets presented here we limited the scan length to 5 cm of OPD (a spectral resolution of 0.2 cm$^{-1}$). With a 30 cm total travel available, this allowed for a maximum baseline change of 15 cm (half of 30 cm spatial change) in the ZPD position without loss of spectral path coverage. The fold mirror scan speed in the spectral arm was limited by the post-detection low-pass filter of the lock-in amplifier (see below) which was set at 300 ms (0.8Hz bandwidth) to average over several chopper cycles. With the additional requirement to wait at least three time constants between samples, this resulted in a total scan time of ~30 minutes per interferogram. Single sided interferograms were recorded, but with sufficient coverage near ZPD to enable a Forman non-linear phase correction to be applied [14]. This correction is only useful for the single slit data taken at short baselines since it removes baseline slopes in the spectral envelope which is needed later for normalization. To first order, this envelope is not dependent on a determination of absolute phase as this serves only to position the spatial fringes in the sky whereas our analysis positions the slit(s) on axis.

Signals at Port 1 and Port 2 at the output of the beam combiner are complementary, being 180$^{o}$ out of phase; the detector is placed at Port 1 which gives a minimum signal at the ZPD. The detector is coupled to the output of the interferometer by an f/4 Winston cone ([15]) with a 10 mm entrance aperture. To limit the high frequency response, multiple metal mesh filters inside the cryostat are used which provide an edge cutoff near 35cm$^{-1}$ and complete rejection of NIR and visible light. Low frequency response is limited by the cut-off wavelength of the 2-mm diameter circular exit aperture of the Winston cone, effectively rejecting frequencies < 2.5 cm$^{-1}$. The detector is a single channel silicon bolometer, operating at a temperature of 4.2 K in a standard QMCIL cryostat (model TK1813). The bolometer is AC coupled to a standard QMCIL low-noise preamplifier (model ULN 9510) mounted on the outside wall of the cryostat with the signal fed into a Stanford Research Systems amplifier/filter bank (model SR650) to band-limit excess noise. Demodulation is achieved using a Stanford Research Systems lockin amplifier (model SR810) which acquires its TTL reference signal from the taut-band modulator drive unit. The analogue output from the lock-in is digitized with a National Instruments PCI-MIO-16XE-IO ADC card which is triggered from the ThorLab translation stage motion controlled via a Labview program. Figure 1 includes a block diagram of the circuitry and Figure 3 shows representative interferograms for the single and double slit configurations.

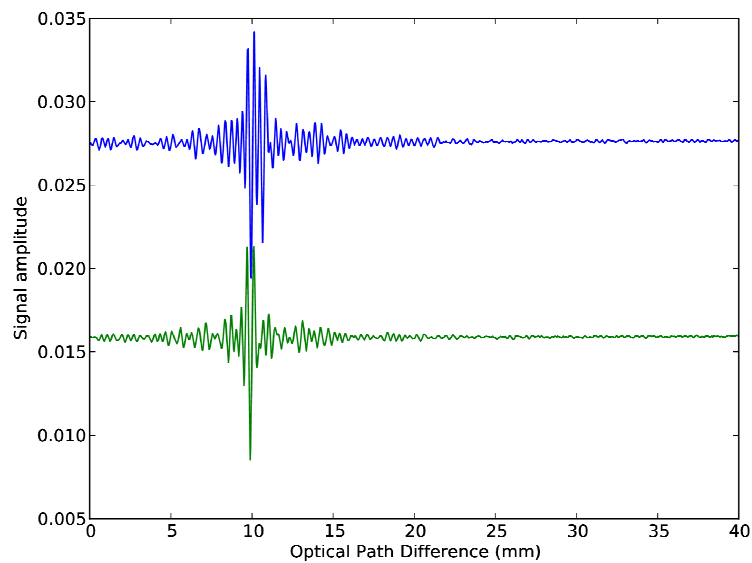

**Figure 3: Representative interferograms for the single (lower) and double (upper) slit data sets**

# 4 Optical alignment and characterization of a single port configuration

The alignment of any interferometer is critical. Because the alignment of the entire system has many variables it was decided to start by only using a single telescope unit to feed both arms of the interferometer. This eliminated critical alignments within the collimation unit and parallelism of the telescope units. The optical configuration was changed by inserting a polarizer at $45^o$ at the stationary telescope output and removing the second telescope unit (spatial arm in Figure 1). Initial alignment was then performed by placing a laser in the final optics directed back towards the Hg arc lamp. The beam divider was replaced with a stretched Mylar beam splitter. Care was then taken to align the optical components such that the beam was at the center of all components and that the beams in the two arms of the interferometer were in the same plane (parallel to the optical bench).

Following the optical adjustments, the THz beam divider was replaced and the laser removed from the beam. The detector was then positioned to maximise the signal from the source with the interferometer set away from the mechanically measured ZPD position. The interferometer should now be close to, but not exactly at, its optimal alignment since the THz beam divider was removed. Final alignment is now performed by finding the ZPD and adjusting the beam divider plane and the interferometer path to minimize the signal (recall we use the negative going port - Port 1 output shown in Figure 1).

Figure 4 shows the spectrum measured in this configuration. The spectral envelope seen whilst viewing the source rises steeply from the detector cut-on as expected from the Rayleigh-Jeans tail of its blackbody-like emission and then falls sharply near 33 $cm^{-1}$ where the beam divider cuts off. The large dips are absorption from the strong 18.58, 25.44, and 32.9 $cm^{-1}$ water vapor lines. Using a model for atmospheric transmission derived from the HiTran spectral line database for the 8 m path length from the source to the detector, we estimate a 15% reduction in power over this frequency band.

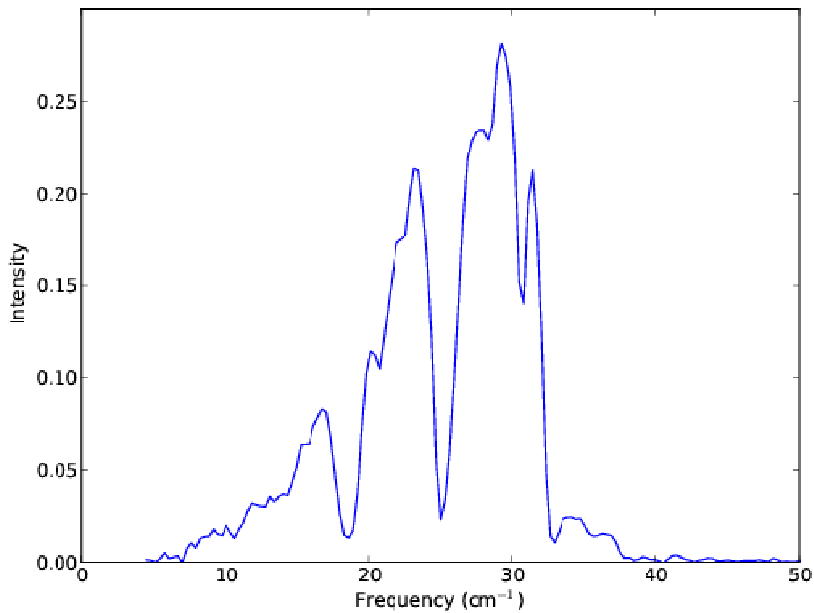

**Figure 4: Low resolution spectrum (resolution = 0.6 $cm^{-1}$) taken with the system as a single input port FTS.**

# 5  Characterization of a dual port configuration

The interferometer was then configured to use both input telescope units as shown in Figure 1 with the polarizer removed. The laser was used, as described above, and the parallelism of the output from both telescope units was ensured, as well as checking the position of all optical elements. We also verified that the source is at the collimator focus by placing it at the point where the two laser beams are coincident. A last check was to scan the positional arm of the interferometer to check that the spot convergence did not walk as the spatial baseline changed. Finally, the THz beam divider was reinstalled, and the final alignment achieved as before.

Figure 5 shows measurements of the vignetting from the spectrometer arm. Ideally an on-axis beam is not affected by the motion of the FTS mirrors since there is no divergence as the mirror separation increases. However, for a finite source size there are non-parallel beams from the off-axis positions which tend to walk-off the interferometer mirrors as the optical path increases. This is clearly seen in Figure 5, where the spatial arm of the interferometer was blocked with a piece of Eccosorb AN72 (16), so no interference fringes were produced. A 38% variation is seen along the length of the spectral arm with the highest throughput occurring when the optical path is at a minimum. Removal of this effect in subsequent analysis is not actually necessary since in the Fourier domain it appears at very low frequency (~ DC). However, the presence of this vignetting does show that in an optimized instrument we need to improve the beam control by using powered FTS mirrors to create a beam waist halfway down the mirror travel to minimize walk-off of any off-axis field rays.

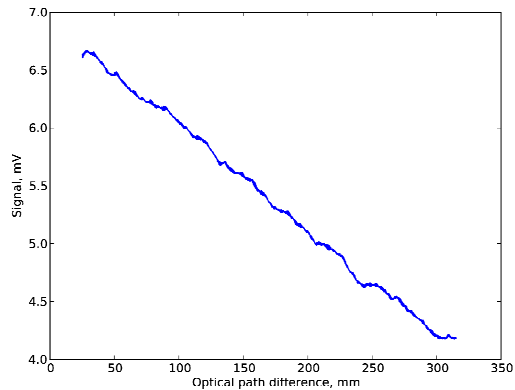

**Figure 5: Vignetting effect of the spectral arm when the interferometer is blocked with an absorber.**

We estimate that the power at the detector window is 0.65 nW, after correcting for atmospheric loss, modulation loss from the chopper and FTS, and the efficiency of the optical components. Combining this with the known detector system NEP (referred to power at the window) of 2.8 pW $Hz^{-1/2}$ gives an achievable instantaneous S/N of 230 which is sufficient for alignment and component evaluation. By differencing the forward and backward interferograms we have determined an instantaneous S/N ratio of 182 from the single slit source data. This indicates that the overall interferometer optical efficiency is about 78%. The main part of this, as seen in Figure 5, comes from the off-axis beam walk-off. In a real system all of these losses would need to be minimized to enable good spectral and spatial resolution of weak astronomical sources.

A subset of single slit spectra for a range of spatial baselines is shown in Figure 6. For each baseline we determine from the spectral interferometric data the local position of the ZPD, and discard information about its absolute position. Recording of the absolute spatial and spectral positions with reference to the phase-center of the double interferometer would allow the source position to be determined. Inspection of Figure 6 shows that as the baseline is increased we see a spatial fringe entering at the highest frequency end as we start to spatially resolve the single slit.

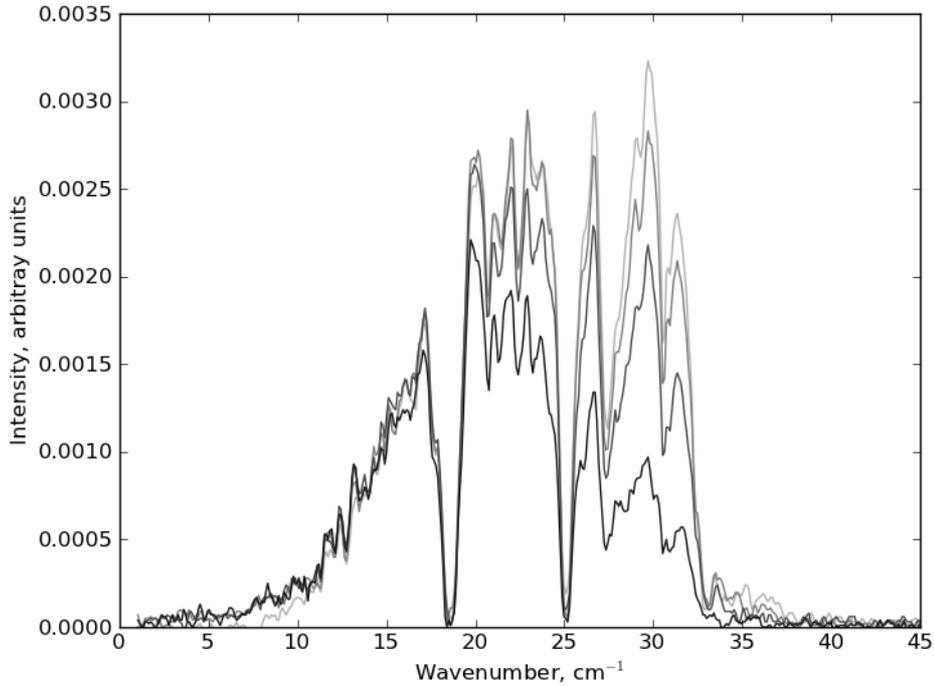

**Figure 6: Spectra for four spatial baselines with a single slit (1.1 mm wide) source. The baseline is indicated by the grayscale; the darkest line is for a baseline 392 mm, then, with successively light lines; 317, 242 and 167mm.**

Equation (7) gives a model for a single slit source. So that we can compare the data with the model, we must first remove the intrinsic intensity variation with wavelength from the source. Given the intensity data (as a function of wavenumber and baseline), we find the maximum intensity for each wavenumber, and use the resulting value as a normalizing factor. This is reasonable as the physical shape of the source, i.e. its width, should have no wavelength dependence in this case. Thus plotting the normalized intensity against baseline, measured in wavelength units ($b/\lambda$), for individual frequencies should give a curve fitting with the model visibility from equation (7). The resulting data and the model are shown in Figure 7. The best fitting slit width is 1.26 mm. As Figure 7 shows, the scatter is high for short baselines since these points are only accessible at the lowest spectral frequencies where the source power is low. At longer baselines the higher spatial frequency data shows less scatter between the different frequency plots. The fit is produced by determining the chi-squared between the model and the data, including an overall scale factor to vary as well as the slit width. Minimization is done with the Nelder-Mead simplex algorithm. We note that the measured slit width is $1.14^{+0.14}_{-0.20}$ mm.

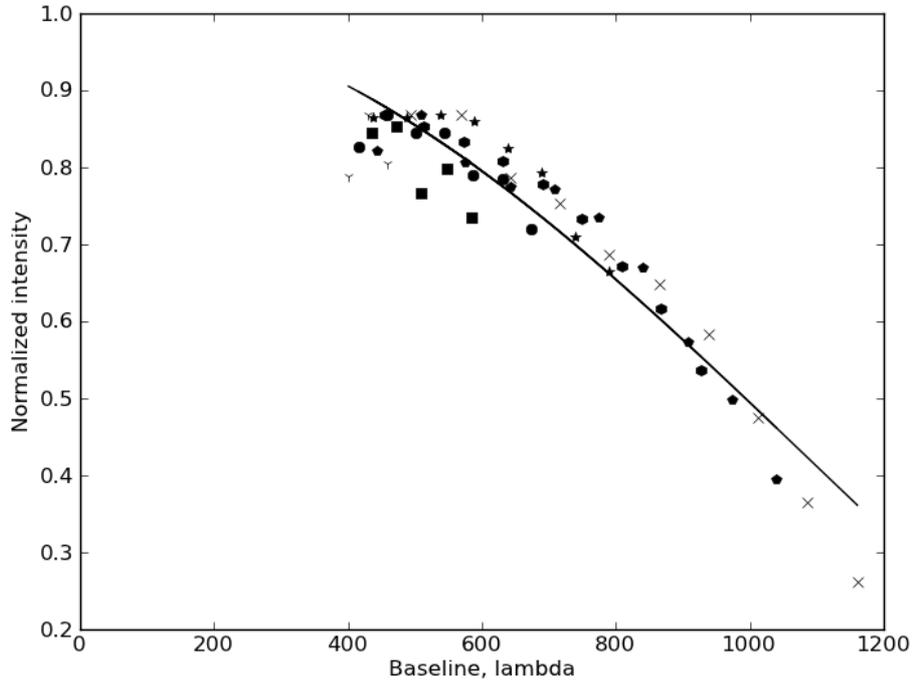

**Figure 7. Solid line shows best fitting model intensity for a single slit source of width 1.26 mm as a function of baseline. Points show the normalized spatial fringe intensity extracted from the spatial/spectral data set at the following frequencies (in wavenumber): 11.7 (three pointed star), 15.0 (square), 17.2 (circle), 20.2 (star), 23.7 (hexagon), 26.6 (pentagon) and 29.7 cm$^{-1}$ (x).**

The next test was to observe a double slit source, as described above, which could be resolved over the baseline range available. Figure 8 shows a subset of the spectral data obtained, analyzed in the same spectral FT method. Again, the absolute position of the sources is lost but the relative phase shift between the positions of the two sources now clearly creates spatial fringes as the baseline changes. A simple interpretation of this experiment is to liken it to a Young's slit experiment with finite apertures but with the additional requirement that we need to add fringe intensities rather than amplitudes as the two slits are incoherent. A better method is to use the formalism of Section 2, recalling that the scene information resides in the complex degree of coherence, $\gamma_{12}$. Following Van Cittert and Zernike, $\gamma_{12}$ should now represent the Fourier transform of a pair of delta functions, convolved with a top hat to represent the slits. Thus we expect to see an intensity variation at the interferometer output given by:

$$I(\phi,b) = \left(\frac{2J_1(u)}{u}\right)^2 \left(1 + |\text{sinc}(v)\cos(v_{double})|\cos(\beta_{12}(v) - \delta)\right), \quad (16)$$

where all the parameters are defined in Section 2.

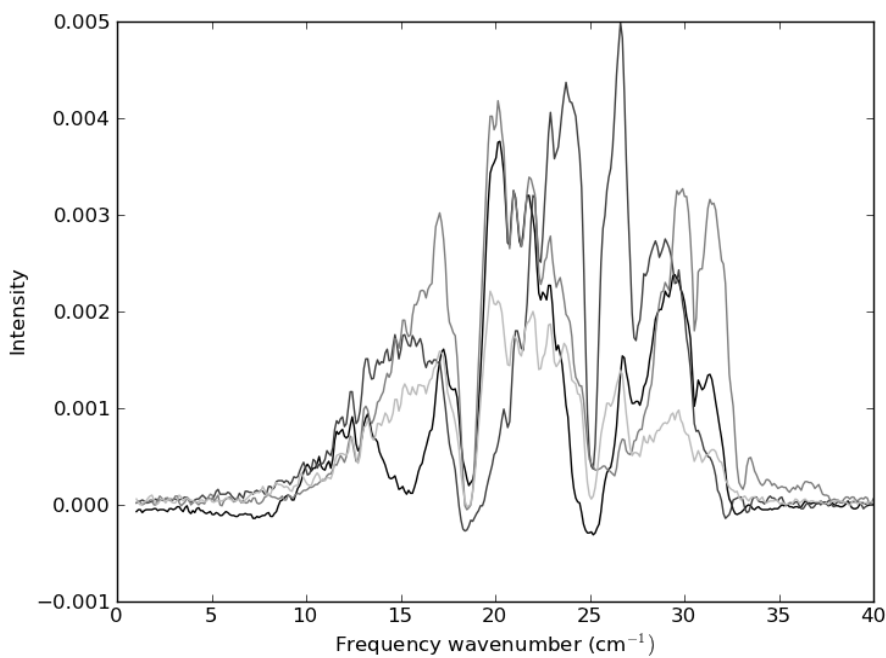

**Figure 8: Spectra for different spatial baselines with a double slit source (two ~1 mm wide slits, separated by 5.5 mm). The baseline is shown by the greyscale; the darkest line is for a baseline 392 mm, then, with lighter lines, 317, 242 and 167mm.**

To normalize the spectra and remove the spectral throughput variations we assume that the two slits are identical, and use the single slit data, multiplied by 2, as our reference. This is justified as Figure 3 shows that the intensity is nearly doubled for the double slit case in comparison to the single slit case. As before, we assume the source shape does not vary with frequency, so we follow the same procedure to normalize the intensities and determine the source parameters. For two monochromatic sources of the same intensity, the expected spatial fringes are *cosinusoidal* within the *sinc* envelope produced from the silt widths. The smallest spatial frequency observable is set by the separation of the sources and the overall spatial phase is set by the distance of the source from the nominal phase centre of the interferometer. Our basic analysis method loses this latter information but we can determine the separation of the two sources which only depends on the period of the fringes seen when plotted against baseline, as shown in Figure 9. Each data set here shows the fringe variation with baseline for a specific spectral frequency and is equivalent to the output expected from a single frequency radiometric interferometer system. As in Figure 7, the shortest baselines use data at the lowest spectral frequencies and thus have lower source intensity and poorer signal to noise ratios. Some of the scatter between different frequencies is also due to the calibration dependency on a single spectral point in the single slit data set. Clearly spectral averaging over say a 1cm$^{-1}$ band (20 spectral points) would lower the scatter and improve the signal to noise ratio in band averaged fringe data sets. Here we have chosen to show the raw data to highlight the noise dependencies. Using equation (16) we have overlaid the best fitting theoretical curve for the double slit apertures in Figure 9 (plotted as $I_{max} - I_{min}$). The same fitting procedure is used as for the single slit case, but with the additional parameter of the slit separation (the slits are assumed to have the same width). The best fit slit width and separation are 1.3 mm and 5.4 mm, respectively. We estimate that the distribution of these is ±0.1mm for slit width and ±0.05mm for the slit separation. This estimation is performed by allowing the variables to vary and observing when the fit looks unreasonable. The position of the nulls, which are insensitive to spectral phase or calibration errors, are a sensitive function of slit separation, leading to a relatively tight constraint on this parameter.

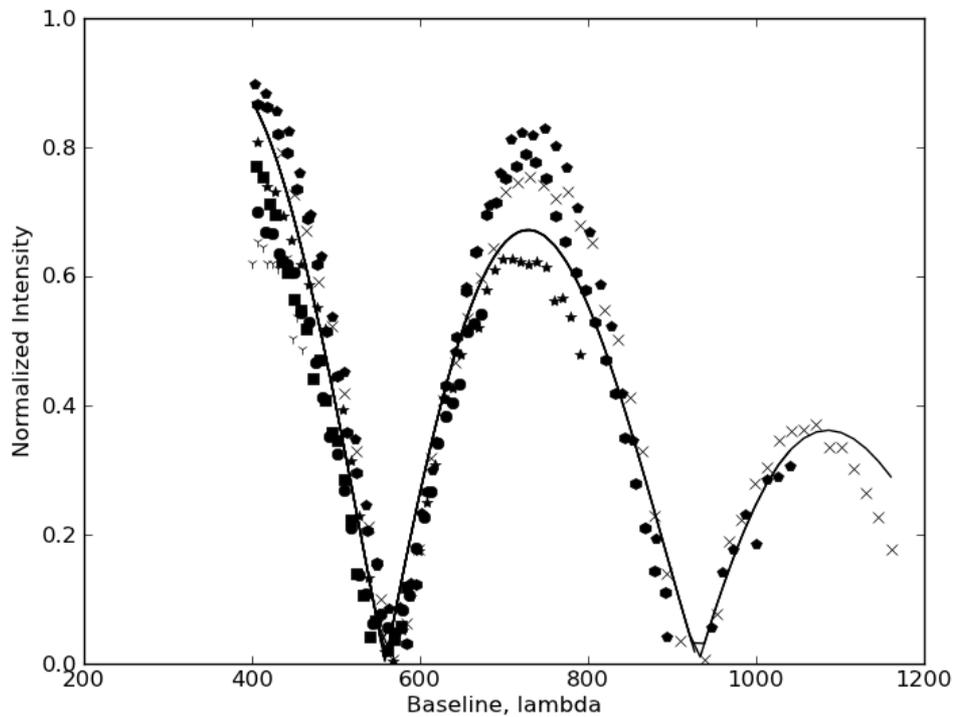

**Figure 9: Best fitting model (line) and data for different frequencies for the double unobstructed slit dataset. The frequencies used and their symbols match those in Figure 7. The best fitting model curve is for a slit width of 1.3mm and a separation of 5.4mm.**

As a last test we covered one of the slits with a 21 cm$^{-1}$ low pass filter to present two sources of different spectral composition and different intensities to the interferometer. The spatial frequencies data are extracted as described above and are shown in the upper left panel of Figure 10. On the upper right panel is the modeled intensity. The frequencies are chosen to cross the edge of the filter transmission, as indicated in the lower panel of Figure 10. The model qualitatively illustrates the shift from the intensity pattern due to a double slit source to that of a single slit source in reasonable agreement with the data. Clearly a more accurate representation of the scene is required to improve this fit.

We note that Figure 10 points to a number of differences between the model and the data. The "null", which is sharp and at a constant frequency in the model, varies slightly with frequency, and is relatively broad in the data. Additionally, any attempt to fit this data shows a gentle slope in the derived filter transmission curve rather than the sharp slope seen in (independent) measurement of the filter.

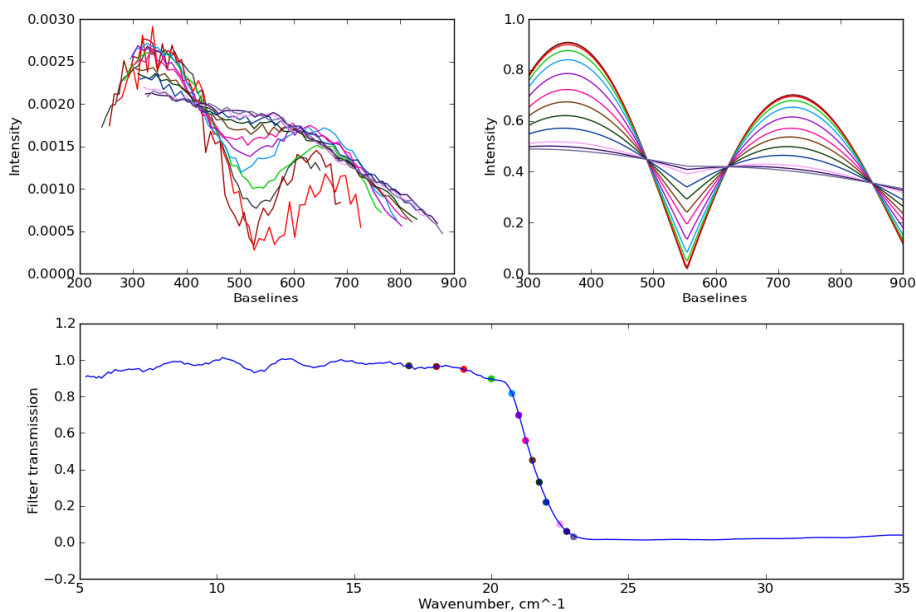

**Figure 10. Double slit plus low pass filter dataset. Upper left panel: data for different frequencies, scaled such that the difference between different sets is minimized. Frequencies used are indicated by the points on the lower plot. Upper right: theoretical intensity curves, again with the same frequencies used. Transmission values are taken from the independently measured transmission curve (lower plot)**

## 6 Discussion and future work

In this article we have described a prototype double Fourier interferometer operating at THz frequencies demonstrating both spectral and spatial capability. We have performed instrument characterization by observing a single slit source, a double slit source and a differentially filtered double source. Since the experimental source configuration is known we have been able to compare the measurements with modeled results. We find that the source geometry and the model fit to the single slit and double slit data agree within our estimated errors; the fitted slit width of 1.3±0.1 mm should be compared with actual widths of 1.14 and 1.01 mm, and a slit separation was found from fitting to be 5.4±0.05mm, and the measured separation is 5.5±0.1mm.

There are still a number of features that can be improved in the system, which can be addressed both with improved modeling and hardware changes. Firstly, the beam walk-off in the spectral arm of the interferometer, as illustrated in Figure 5, should be reduced by using beam focusing optics. Secondly, the mechanical construction of the aperture mask is not robust, and if we are to move to higher frequencies then the uniformity of the slit edges and parallelism need to be more carefully controlled by utilizing photolithographic made aperture masks. In further work, more robust scenes will be used. Slits were used to increase the available signal, but the height of the slit is not considered in our simple mathematical model. Thirdly, our model does not account for any possible misalignment in the collimator. Finally, there exists the possibility of unbalanced spectral throughput in each interferometer arm which can lead to spectral phase errors. We can check this by moving the detector to port 2.

In our data processing, we did not record the absolute position of the FTS scanning mirror as the baseline was changed and hence we lost the positional information of the source with respect to the interferometer axes. We also used a step and sample scan method which is inefficient compared to rapid scan techniques for the spectral drive. A future upgrade is thus to install a rapid scan drive with improved metrology to increase the overall observing efficiency and provide absolute phase measurement.

As we explore more complex scenes more sophisticated data processing techniques must be considered as used in radio astronomy interferometry[17]. These include observations of known point like sources to calibrate the instrument or taking advantage of techniques such as self-calibration, where a known point source in a complex scene is used to calibrate the system.

In this paper, we have reported on a single data taking technique; scanning the spectral arm and then stepping the spatial arm. We also intend to investigate the reverse: scanning the spatial arm and stepping the spectral arm. Although nominally identical, they should be compared to inform future mission design studies. In addition, we intend to extend the testbed to work at shorter wavelengths, allowing for direct evaluation and optimization of BETTII observing strategies and data processing techniques.

## Acknowledgements

We are grateful to the BLAST team for the loan of their mirror segment. Work in Cardiff is supported by a STFC rolling grant. Work in UCL is supported by Perren Foundation and IMPACT scheme. We are grateful to the engineering support team in Cardiff.